\documentstyle{mn}

\begin{document}
\title {X-ray properties of the microquasar GRS~1915$+$105 during a variability class transition}
\author[Naik, S. et al.]
{S. Naik,  P. C. Agrawal, A. R. Rao, and B. Paul \\
Tata Institute of Fundamental Research, Homi Bhabha Road, Mumbai, India 400~005\\
sachi@tifr.res.in (SN), pagrawal@tifr.res.in (PCA), arrao@tifr.res.in (ARR), bpaul@tifr.res.in (BP)}
\date{Accepted for publication in MNRAS, 2001}
\maketitle
\begin{abstract} 
We present a detailed X-ray study of the microquasar
GRS~1915$+$105 during a variability class transition observed in 2000 June 
with the Pointed Proportional Counters (PPCs) of the Indian X-ray Astronomy 
Experiment (IXAE). We supplement this observation with data from the 
RXTE archives. The source made a transition from a steady low-hard state 
to a regular oscillatory behaviour in the light curve known as bursts or 
class $\rho$ (Belloni et al. 2000) between 2000 May 11 and 17 and reverted 
back to the low-hard state on 2000 June 27. A gradual change in the burst 
recurrence time from about 75 s to about 40 s was observed which then 
increased to about 120 s during the $\sim$ 40 days of class $\rho$. The 
regular bursts disappeared from the X-ray light curves and the class 
transition was observed to occur within 1.5 hours on 2000 June 27 with 
the PPCs. A correlation is found between the observed QPO frequency at 
5$-$8 Hz in the quiescent phase and the average X-ray intensity of the 
source during the class $\rho$. We notice a strong similarity between the 
properties of the source during the class $\rho$ and those during the 
oscillatory phase of the observations of class $\alpha$ which contains 
a long stretch ($\sim$ 1000 s) of steady low-hard state in the light curve 
along with the regular periodic bursts. From the timing and spectral 
analysis, it is found that the observed properties of the source over 
tens of days during the class $\rho$ are identical to those over a time 
scale of a few hundreds of seconds in the class $\alpha$. Examining the 
light curves from the beginning of the RXTE/PCA and RXTE/ASM observations, 
it is found that the change of state from radio-quiet low-hard state to 
high state occurs through the X-ray classes $\rho$ and $\alpha$ which 
appear together during the state transition. It is further inferred that 
the source switches from low-hard state to the class $\rho$ through the 
intermediate class $\alpha$.

\end{abstract}


\section{Introduction}
The Galactic X-ray source GRS~1915$+$105, discovered in 1992, was 
identified with a superluminal radio source at a distance of 12.5 $\pm$
1.5 kpc (Mirabel \& Rodriguez 1994). Its  radio characteristics like jets
and superluminal motion are similar to those found in quasars and hence
this source is called a microquasar. The source is very bright in X-rays and
shows strong variability over a wide range of time scales. The X-ray
emission is characterized by quasi-periodic oscillations (QPOs) at centroid
frequencies  in the range of 0.001 $-$ 67 Hz (Morgan, Remillard,
\& Greiner 1997). Based on extensive X-ray studies, Muno, Morgan, \& Remillard
(1999) classified the behavior of the source into two distinct states,
a spectrally hard-state with the presence of narrow QPOs, dominated by a
power-law component and a soft-state with the absence of QPOs, dominated by
thermal emission.

Belloni et al. (2000) have made an extensive study of
the X-ray emission of the source and classified all the publicly
available RXTE/PCA observations from 1996 January to 1997 December
into 12 different classes on the basis of structure of the X-ray
light curve and the nature of the color-color diagram. They found that
the source variability is restricted into three basic states, a
low-hard state with invisible inner accretion disk (C),
a high-soft state with visible inner accretion disk (B) and a low-soft state
with spectrum similar to the high-soft state but with much less intensity (A).
However, in GRS 1915$+$105 the observed fast and slow transitions from
one state to other are not clearly understood.

The different variability classes of Belloni et al. (2000) range from
steady emission for long durations like class $\phi$ (state A), class $\chi$
(state C) to large amplitude variations (from state C to state B with
a hint of state A) in classes $\lambda$, $\kappa$, $\rho$. In the high
state (state B), the source sometimes shows short periodic flickering
with different amplitudes (classes $\gamma$, $\mu$ and $\delta$). During
the classes $\theta$, $\beta$ and $\nu$ the amplitude variation is
accompanied by soft X-ray dips (state A) with duration of a few
tens of seconds to hundreds of seconds. It has been suggested that these 
soft dips in X-ray light curves are also associated with radio flares 
(Naik et al. 2001). The class $\chi$ is further divided into four sub-classes 
$\chi1$, $\chi2$, $\chi3$ and $\chi4$, depending on the count rate and hardness
ratio in different energy ranges. During the class $\rho$, the X-ray light
curve consists of regular and characteristic pattern known as bursts (Taam,
Chen, \& Swank 1997; Yadav et al. 1999)  which is reflected
as a loop-like behaviour (Vilhu \& Nevalainen 1998) in the color-color diagram.
Similar type of oscillations are observed in the X-ray light curve for a
duration of a few hundreds of seconds followed by a long quiet period of about
1000 s as seen in the class $\chi2$/$\chi4$. The observations with these 
characteristics are as classified into class $\alpha$. Among the twelve 
different classes of X-ray observations, it is seen that the source is 
radio-loud in X-ray classes $\theta$, $\beta$, $\chi1$ and $\chi3$ (Naik \& Rao
2000).

We present the results of the observation of GRS~1915$+$105 with the 
PPCs onboard the IXAE when the source made a class transition from a regular
oscillatory behaviour of class $\rho$ to a steady state of class 
$\chi2$/$\chi4$. We have also analyzed the available RXTE archival data 
contemporaneous to the PPC observation. In the following sections we present
results of these studies.

\section{Instrument and Observations}

The X-ray observations of the microquasar GRS 1915$+$105 were
made using the PPCs of the IXAE on board the Indian satellite
IRS-P3. The IXAE includes three co-aligned, identical, multi-wire,
multi-layer proportional counters with an effective collecting area
of 1200 cm$^{2}$ and a field of view of 2.3$^{\circ}$ $\times$ 2.3$^{\circ}$.
All the PPCs operate in the energy range of 2${-}$18 keV with an average
detection efficiency of about 60\% in 3$-$7 keV energy range.
For a detailed description of the PPCs refer to Agrawal (1998)
and Rao et al. (1998). Background observations were made before and
after the source observation by pointing the PPCs to a source-free
region in the sky, close to the target source. The source GRS 1915$+$105
was observed from 2000 June 18 to 22 with 1 s time integration mode and
from June 23 to 27 in 0.1 s integration mode for a total useful period of
29460 s. The log of the PPC observations is given in Table 1.

Regular pointed observations of the microquasar GRS~1915$+$105 during 1996 $-$
2001 period with the Rossi X-ray Timing Explorer (RXTE) satellite provide a good
coverage for a detailed study of the timing and spectral properties of the 
source and establish the associated radio properties with X-ray emission during
different spectral states. There have been over 550 observations of 
GRS~1915$+$105 with the PCA and HEXTE of the RXTE. The list of observations 
used in the present work is given in Table 2.

\begin{figure}
\vskip 12.4cm
\includegraphics{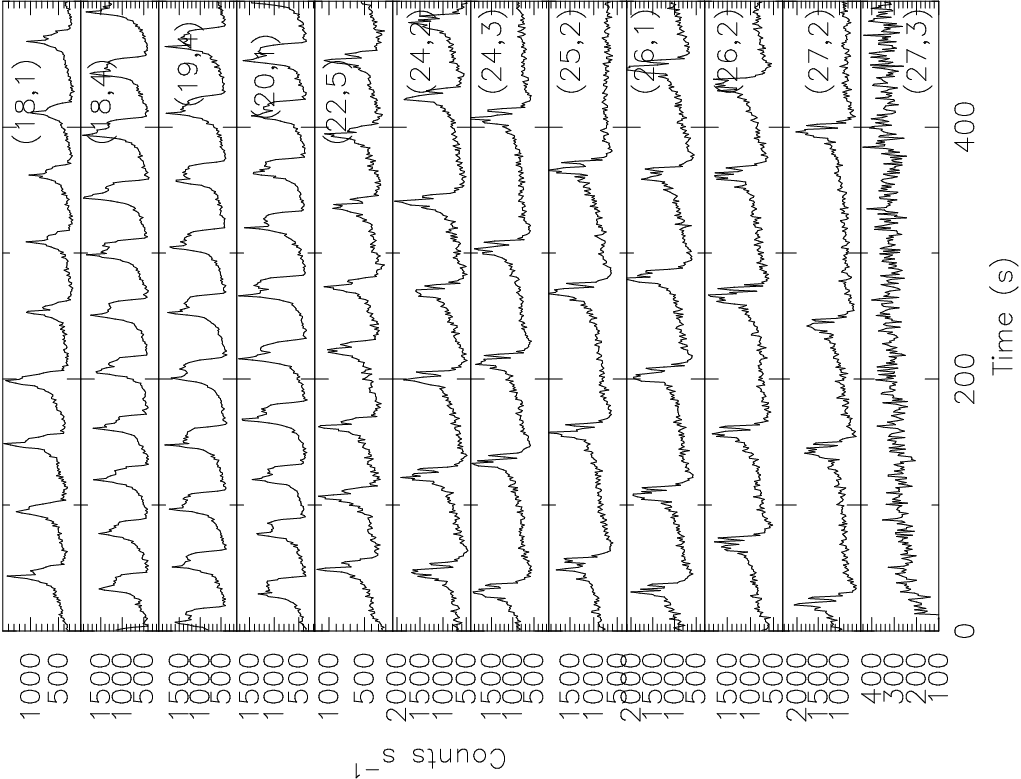}
\caption[fig1.eps]{The X-ray light curves for GRS~1915$+$105 obtained with the
PPCs (PPC$-$1 and PPC$-$3 added) in the energy range 2$-$18 keV in 1 s
time integration mode from the observations in
2000 June. The date and orbit of the observations are indicated in each
panel of the figure.}
\end{figure}

\section{Analysis and Results}
\subsection{Data from IXAE Observations}

The X-ray data in the energy band 2$-$18 keV and 2$-$6 keV were
corrected for pointing offset using the aspect information. The
background count rates obtained by pointing the detectors to a
source-free region, were subtracted from the source count rate. Dead time 
correction has been neglected as it is less than 1\% even at a count rate
of about 500 counts s$^{-1}$ per PPC. From calibration of the PPCs using 
Crab Nebula, it is found that the spectral data from PPC$-$3 is more 
reliable. Hence the hardness ratio (ratio of the count rates in 6$-$18 
keV and 2$-$6 keV bands) has been obtained using data only from PPC$-$3.

\begin{figure*}
\vskip 13 cm
\includegraphics{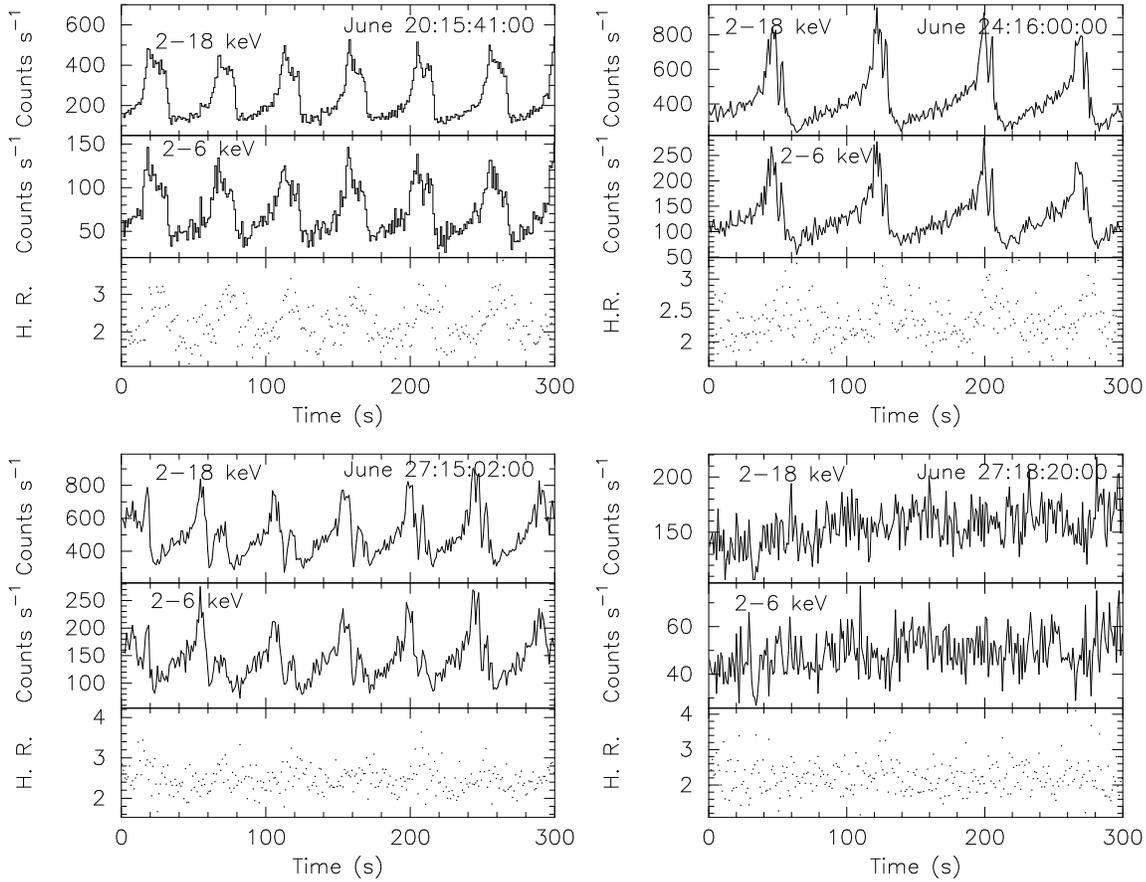}
\caption{The X-ray light curves of GRS~1915$+$105 obtained from
PPC$-$3 of IXAE during four different intervals of 2000 June observation
are shown in the energy ranges 2$-$18 keV and 2$-$6 keV along with hardness
ratio H. R. (ratio of count rate in 6 $-$ 18 keV range to count rate in
2$-$6 keV range). The appearance of secondary (and tertiary peaks) can be
seen in the later part of the observation.}
\end{figure*}

We have generated the X-ray light curves of the source in 2$-$18 keV range 
for 500 s using the data from PPC$-$1 and PPC$-$3 for each orbit of 
observation. The light curves for a few orbits are shown in Figure 1. 
The light curves show that low intensity quiescent phase lasting for a few 
tens of seconds, is followed by the bursts, which have exponential increase
in X-ray flux with rise time of $\sim$ 8$-$10 s, high peak flux level
for $\sim$ 10 s and a sharp linear decay in $\sim$ 2$-$3 s.
The average burst recurrence time (the time taken by a cycle of quiescent 
interval followed by a burst) for all the orbits of the PPC observations 
and the burst strength (the ratio between the peak count rate of the burst 
and the quiescent count rate at 2$-$18 keV for PPC$-$1 and PPC$-$3) are given
in Table 1. An increase in the burst recurrence time by a factor of about 2 
from the beginning to the end of the observations is clearly detectable from
the table. The X-ray light curves in 2$-$18 keV and 2$-$6 keV bands along 
with the hardness ratio obtained from PPC$-$3 data for four different 
PPC observations are shown in Figure 2. It can be seen that the spectrum 
becomes hard as the burst progresses and it is hardest at the end of the 
decay of the burst. Presence of secondary (and tertiary) peaks during the 
decay phase of the bursts in the X-ray light curves is also discernible.
From the light curves and hardness ratio for the third orbit of PPC 
observations on June 27 shown in the fourth panel of Figure 2, it can be 
seen that the regular bursts disappeared in both the energy bands and the 
transition from class $\rho$ to class $\chi2$/$\chi4$ occurred within about 
1.5 hours, the orbital period of IRS-P3 satellite.

\begin{table}
\centering
\caption{Log of X-ray Observation of GRS~1915$+$105 with the PPCs of IXAE}
\begin{flushleft}
\begin{tabular}{lllllll}
\hline
\hline
\multicolumn{2}{|c|}{Observation} &Start time &End Time  &BRT$^2$ &Burst\\
Date$^1$  &orbit  &(UT)  &(UT)  &(s) &strength$^3$\\
\hline
\hline
18    &1    &13:00   &13:18     &53       &2.45$\pm$0.15\\
      &4    &18:04   &18:23     &48       &2.87$\pm$0.16\\
19    &1    &12:44   &12:54     &46       &2.95$\pm$0.16\\
      &2    &14:20   &14:38     &46       &2.79$\pm$0.14\\
      &3    &16:02   &16:20     &50       &3.14$\pm$0.17\\
      &4    &17:43   &18:02     &51       &2.87$\pm$0.21\\
      &5    &19:35   &19:43     &47       &2.93$\pm$0.15 \\
20    &1    &12:16   &12:36     &47       &3.24$\pm$0.17\\
      &2    &14:00   &14:16     &49       &2.81$\pm$0.16\\
      &3    &15:41   &16:00     &48       &3.42$\pm$0.20 \\
      &4    &17:23   &17:42     &46       &3.09$\pm$0.19 \\
      &5    &19:12   &19:22     &49       &3.40$\pm$0.17 \\
21    &2    &13:42   &13:57     &43       &3.19$\pm$0.18\\
      &3    &15:21   &15:38     &48       &3.55$\pm$0.19 \\
      &4    &17:03   &17:21     &53       &3.57$\pm$0.22 \\
      &5    &18:49   &19:02     &52       &3.33$\pm$0.18 \\
22    &4    &16:41   &16:59     &54       &3.30$\pm$0.18\\
      &5    &18:27   &18:40     &57       &2.55$\pm$0.16 \\
23    &2    &16:21   &16:37     &49       &2.80$\pm$0.18 \\
      &3    &18:00   &18:19     &49       &2.94$\pm$0.28  \\
24    &1    &14:25   &14:34     &77       &3.60$\pm$0.19 \\
      &2    &15:59   &16:16     &75       &2.89$\pm$0.14 \\    
      &3    &17:39   &17:58     &93       &2.61$\pm$0.13  \\
25    &1    &14:07   &14:13     &85       &3.02$\pm$0.13 \\
      &2    &15:37   &15:54     &107      &2.75$\pm$0.12 \\
      &3    &17:17   &17:36     &92       &2.49$\pm$0.11 \\
26    &1    &15:17   &15:34     &80       &2.67$\pm$0.12 \\
      &2    &16:58   &17:16     &98       &2.51$\pm$0.12 \\
27    &1    &15:01   &15:12     &54       &1.77$\pm$0.08 \\
      &2    &16:36   &16:54     &126      &2.34$\pm$0.12 \\
      &3    &18:18   &18:35     &$----$   &$----$        \\ 
\hline
\hline
\end{tabular}
\end{flushleft}
\begin{flushleft}
$^1$Date $-$ 2000 June, \\
$^2$BRT $-$ Burst Recurrence Time (averaged over the orbit)\\
$^3$Burst strength $-$ Ratio of Peak count rate to quiescent count rate \\
\end{flushleft}
\end{table}

The burst strength has been calculated from a typical burst from each 
PPC observation. The peak count rate was determined as the average count 
rate over a time span of 4 seconds during the peak of the bursts and the 
quiescent count rate as the average count rate over 20 seconds just after 
the decay of a burst. The variation of the burst recurrence time with the 
burst strength is shown in Figure 3. An anti correlation is found between 
the count rate ratio (burst strength) and the average burst recurrence time 
with a correlation coefficient of -0.54 (30 degrees of freedom). This 
indicates that the bursts were stronger during the early phase of observations 
when the burst recurrence time was small and the burst strength decreased 
with the increase in burst recurrence time.  

\begin{figure}
\vskip 7 cm
\includegraphics{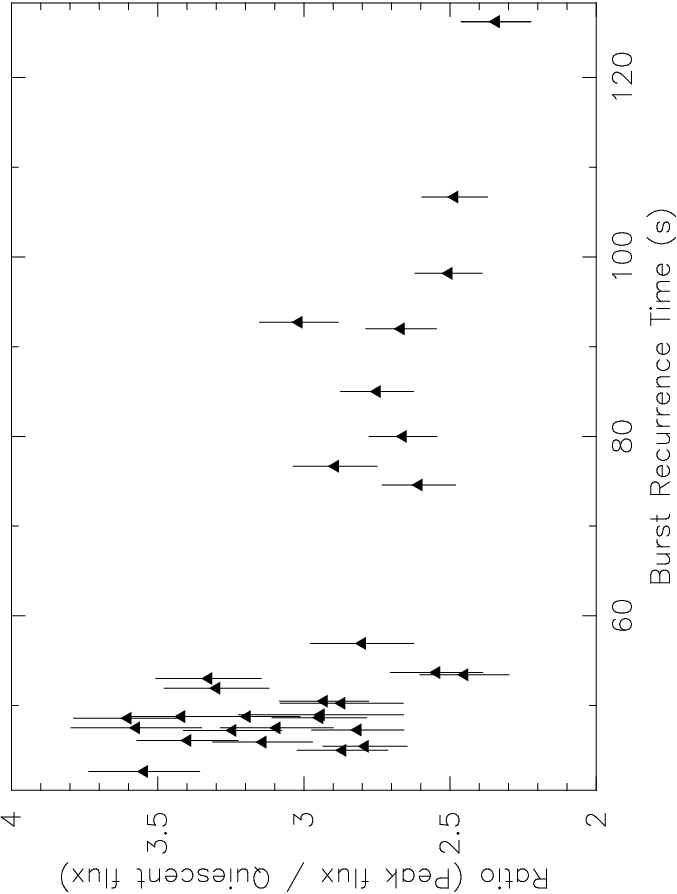}
\caption[fig3.eps]{Plot of the average burst recurrence times obtained from
IXAE versus the ratio of the peak flux and quiescent flux in 2 $-$ 18 keV
energy is shown. A negative correlation between the burst duration and the
ratio between the peak flux and quiescent flux is seen.}
\end{figure}

\subsection{Data from RXTE Observations}

To corroborate the IXAE results we have also analyzed 31 RXTE/PCA observations 
of GRS~1915$+$105 made between 2000 May 4 and 2000 July 5 which are listed
in Table 2. Other details like the average count rate during quiescent phase 
and burst phase, burst recurrence time, the observed low frequency QPOs 
($\sim$ 2$-$8 Hz), hardness ratios and rms variability are also given in 
the table. The hardness ratio HR1 is described as the ratio of the count 
rates in the energy range 5$-$13 keV to that in 2$-$5 keV whereas the ratio 
of the count rates in the energy bands of 13$-$60 keV and 2$-$13 keV is 
described as HR2. The source was in a low-hard state ($\chi2$/$\chi4$) on and 
before 2000 May 11 and changed to the oscillatory nature of class $\rho$ and 
remained for $\geq$ 40 days (from 2000 May 17 to 2000 June 27, as observed 
with RXTE/PCA). It returned back to the low-hard state which was detected 
on 2000 July 5. The burst recurrence time for each of the RXTE observation 
of class $\rho$ was calculated using the PCA X-ray light curves. A plot of 
the average burst recurrence time versus the observation date (MJD) for the 
combined IXAE and RXTE/PCA observations is shown in Figure 4. It can be seen 
that the burst recurrence time, obtained from the RXTE was about 75 s in the 
beginning and it gradually decreased to $\sim$ 40 s. Towards the end of the 
observations, regular monitoring by PPCs revealed a gradual increase in the 
burst recurrence time to $\sim$ 120 s which is followed by the disappearance 
of the bursts from the light curve. This indicates the switching of the source 
from oscillatory behaviour of class $\rho$ to a radio-quiet steady emission of 
class $\chi2$/$\chi4$.

\begin{table*}
\centering
\caption{List of RXTE/PCA observations analyzed from the archival data}
\begin{flushleft}
\begin{tabular}{llccccllllllll}
\hline
\hline
\multicolumn{2}{|c|}{RXTE Observation $^1$} &\multicolumn {3}{|c|}{Average Countrate (2 PCUs)}    &QPO frequency  &BRT$^4$ &\multicolumn {3}{|c|}{Quiescent phase}\\ 
ID   &Date     &QP$^2$  &BP$^3$  &Total  &(in Hz) &(in s) &HR1$^5$  &HR2$^6$ &RMS\\
\hline
\hline
         &  &  &  &class $\rho$ \\
\hline
\hline
50703-01-10-00  &05/17/2000  &3317 &10684 &3822  &6.28$\pm$0.062 &73.68 &0.68  &0.055  &0.068\\
50703-01-10-01  &05/17/2000  &3182 &9651  &3703  &6.67$\pm$0.079 &66.67 &0.66  &0.053  &0.062\\
50703-01-10-02  &05/17/2000  &3088 &9207  &3560  &6.48$\pm$0.076 &66.67 &0.66  &0.059  &0.064\\
50703-01-11-00  &05/25/2000  &2894 &9514  &3466  &6.62$\pm$0.138 &55.56 &0.58  &0.056  &0.068\\
50703-01-11-01  &05/25/2000  &3054 &9264  &3556  &6.44$\pm$0.077 &56.00 &0.57  &0.059  &0.065\\
50703-01-11-03  &05/25/2000  &2724 &8456  &3344  &6.15$\pm$0.087 &50.00 &0.57  &0.058  &0.064\\
50703-01-12-00  &05/31/2000  &2642 &9443  &3320  &6.41$\pm$0.067 &46.67 &0.51  &0.059  &0.056\\
50703-01-12-01  &05/31/2000  &2827 &10104 &3365  &6.42$\pm$0.053 &50.00 &0.52  &0.059  &0.067\\
50703-01-12-02  &05/31/2000  &2672 &10554 &3340  &6.32$\pm$0.064 &47.37 &0.52  &0.056  &0.062\\
50703-01-12-03  &05/31/2000  &2540 &9626  &3183  &6.48$\pm$0.146 &46.15 &0.52  &0.056  &0.063\\
50703-01-13-00  &06/07/2000  &2859 &9415  &3517  &6.74$\pm$0.0515 &42.50 &0.48 &0.053  &0.047\\
50703-01-13-01  &06/08/2000  &2908 &9838  &4703  &6.60$\pm$0.0435 &46.67 &0.50 &0.054  &0.057\\
50703-01-13-02  &06/08/2000  &2652 &8601  &3467  &6.56$\pm$0.0585 &47.06 &0.49 &0.051  &0.063\\
50703-01-13-03  &06/08/2000  &2668 &9668  &3495  &6.68$\pm$0.085  &46.15 &0.49 &0.049  &0.066\\
50703-01-14-00  &06/14/2000  &3875 &9182  &5035  &6.74$\pm$0.0295 &57.14 &0.61 &0.051  &0.066\\
50703-01-14-01  &06/15/2000  &3617 &9523  &4237  &6.83$\pm$0.0435 &54.55 &0.59 &0.051  &0.061\\
50703-01-14-02  &06/15/2000  &3521 &9004  &4140  &6.80$\pm$0.0335 &53.85 &0.59 &0.051  &0.052\\
50703-01-15-00  &06/19/2000  &2777 &8794  &5206  &6.81$\pm$0.045  &46.51 &0.48 &0.048  &0.056\\
50703-01-15-01  &06/20/2000  &3066 &10068 &5301  &6.70$\pm$0.0535 &49.18 &0.48 &0.048  &0.049\\
50703-01-15-02  &06/20/2000  &2967 &10463 &5399  &6.91$\pm$0.06   &46.51 &0.47 &0.046  &0.052\\
50703-01-15-03  &06/20/2000  &2968 &9910  &5235  &6.84$\pm$0.043  &48.72 &0.47 &0.049  &0.054\\
50703-01-16-01  &06/26/2000  &5226 &11574 &6217  &7.22$\pm$0.0305 &80.65 &0.57 &0.046  &0.065\\
50703-01-16-00  &06/26/2000  &5483 &10697 &6332  &7.34$\pm$0.0395 &70.59 &0.58 &0.043  &0.062\\
50703-01-16-02  &06/27/2000  &4941 &10639 &5455  &7.27$\pm$0.082  &72.73 &0.57 &0.047  &0.069\\
50703-01-16-03  &06/27/2000  &5274 &9926  &5717  &7.23$\pm$0.05   &73.33 &0.57 &0.051  &0.065\\
20402-01-03-00 $^{\clubsuit}$  &11/19/1996  &4987 &9863 &5598  &7.46$\pm$0.011 &76.92  &1.03  &0.068   &0.067\\
\hline
\hline
  &  &  &  &class $\alpha$ and $\chi2$\\
\hline
\hline
20402-01-28-00 $^{\clubsuit}$   &05/18/1997 ($\alpha$) &3442 &5480 &------  &------------ &--------  &1.07  &0.085   &0.084\\
20402-01-04-00 $^{\clubsuit}$   &11/28/1996 ($\chi2$)     &-----  &----- &4414  &4.57$\pm$0.016 &-------- &1.16  &0.099  &0.116\\
50703-01-08-00  &05/04/2000 ($\chi2$) &-----  &----- &3119  &3.37$\pm$0.016 &-------- &0.87  &0.084  &0.141\\
50703-01-08-01  &05/04/2000  ($\chi2$) &-----  &----- &3226  &3.61$\pm$0.017 &--------  &0.86  &0.082 &0.137 \\
50703-01-09-00  &05/11/2000  ($\chi2$) &-----  &----- &2977  &3.96$\pm$0.025 &--------  &0.775 &0.078 &0.124 \\
50703-01-09-01  &05/11/2000  ($\chi2$) &-----  &----- &2703  &3.82$\pm$0.0305 &--------   &0.78  &0.082 &0.133\\
50703-01-17-00  &07/05/2000  ($\chi2$) &-----  &----- &3538  &2.97$\pm$0.0125 &--------  &0.752  &0.077 &0.164 \\
50703-01-17-01  &07/05/2000  ($\chi2$) &-----  &----- &3302  &2.78$\pm$0.0125 &--------  &0.766  &0.08 &0.168\\
\hline
\hline
\end{tabular}
\end{flushleft}
\begin{flushleft}
$^1$      Data used for spectral analysis are indicated by $\clubsuit$  \\
$^2$QP : Quiescent phase,  $^3$BP : Burst phase \\
$^4$BRT: Burst recurrence time (averaged over the orbit of observation)\\
$^5$HR1: Ratio of count rate in 5 $-$ 13 keV to count rate in 2 $-$ 5 keV \\
$^6$HR2: Ratio of count rate in 13 $-$ 60 keV to count rate in 2 $-$ 13 keV \\ 
\end{flushleft}
\end{table*}

We have used the 8 ms time resolution PCA data in 2$-$13 keV band
to generate the power density spectra (PDS) for all the RXTE/PCA
observations listed in Table 2. During the low-hard X-ray states 
(on 2000 May 4 and 2000 July 5, class $\chi2$/$\chi4$), QPOs are detected 
at frequencies 2.8$-$3.9 Hz whereas during the rest of the PCA observations 
(class $\rho$) QPOs are observed at higher frequencies (6$-$8 Hz).
A plot of the QPO frequency and the source flux in 2$-$60 keV band for 2
PCUs during the observations of class $\rho$, is shown in Figure 5. A strong  
correlation is found between the QPO frequency and the average X-ray flux 
with a correlation coefficient of 0.85 (for 25 degrees of freedom).

Various X-ray characteristics of GRS~1915$+$105 such as nature of the 
light curve, gradual change in the burst recurrence time, hardness
ratios HR1 and HR2, presence of QPOs and spectral properties in class 
$\rho$ are found to be similar to those in the oscillatory phase of class 
$\alpha$ (Belloni et al. 2000). These properties are described in the 
next section.

\begin{figure}
\vskip 8.3 cm
\includegraphics{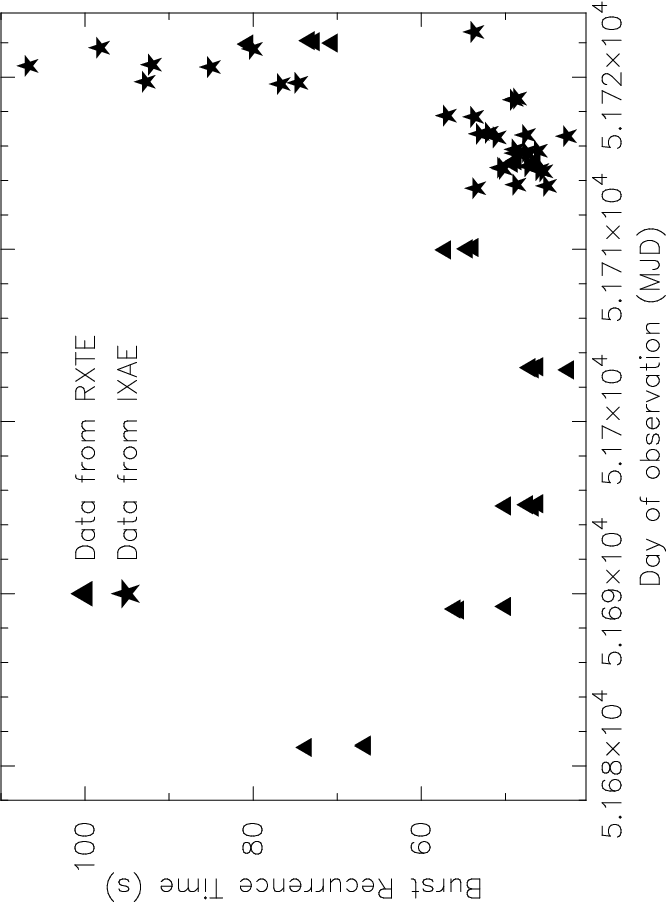}
\caption[fig3.eps]{The average burst recurrence times obtained from IXAE and
RXTE/PCA observations during the long stretch of X-ray class $\rho$ in
2000 May $-$ June are plotted with the day of observation of the source
(MJD). A change in burst duration with observation dates can be clearly seen.}
\end{figure}

\section{Comparison between classes $\rho$~and~$\alpha$}
\subsection{Timing Analysis}

To compare the X-ray timing properties of GRS~1915$+$105 in the X-ray
classes $\rho$ and $\alpha$, we have selected one RXTE/PCA observation
of each class (Obs. IDs 20402-01-03-00 and 20402-01-28-00 respectively) 
when all the 5 PCUs were on. Figure 6 shows the X-ray light curve for the above 
two RXTE/PCA observations for 1 s time bin obtained from the standard-1 
data. The upper panel shows the presence of regular bursts with recurrence 
time of $\sim$ 60 s in the X-ray light curve of class $\rho$  whereas bursts 
with increasing recurrence time and decreasing peak intensity followed by a 
low-hard state lasting for $\sim$ 1000 s (class $\alpha$) are shown in the
bottom panel of Figure 6. At the end of the long stretched ($\sim$ 1000 s)
low-hard state,the bursts again reappear with a very high peak X-ray flux 
(similar to the peak flux of class $\rho$) which decreases gradually with 
increasing burst duration. We have shown, in Figure 7, the X-ray light curve 
for 1 s time bin in 2$-$60 keV energy range along with the hardness ratios 
(HR1 and HR2) for classes $\rho$ (left panels) and $\alpha$ (right panels). 
It is seen that all the properties (structure of the light curves and the 
hardness ratios) are identical during the burst/quiescent phase of the 
two X-ray classes. The presence of QPOs in the PDS during the burst/quiescent 
phase of the classes $\rho$ and $\alpha$ has been shown by Muno et al. (1999) 
(Fig. 1 (d) and 1(j)). From those figures, it can be seen that the QPO 
frequency during the particular observation of class $\rho$ lies between 
5$-$7 Hz. During the class $\alpha$, the QPO frequency changes from $\sim$ 8 
Hz at the beginning of the burst phase, to $\sim$ 5 Hz during the change of 
state to the low-hard state. As the QPO frequency varies directly with the 
intensity, the change in QPO frequency during the burst phase of class 
$\alpha$ is due to a decrease in the X-ray intensity with the increase of 
recurrence time. In both the classes, the source is radio-quiet with similar 
flux densities at 2.25 GHz and 8.3 GHz (Naik and Rao 2000).

From the IXAE and RXTE observations of the source during 2000 May$-$June
(when the source was in the X-ray class $\rho$), it is noticed that 
there is a gradual change in the various properties of the source like 
burst recurrence time, intensity, QPO frequency etc. and the source makes
a gradual transition to the low-hard state of the class $\chi2$/$\chi4$.
These characteristics are very similar to those observed in the 
observations of class $\alpha$ but at a much shorter time scale. Hence, 
we argue that the nature of the source over time scale of a few tens of days 
during the class $\rho$ is identical to that over a time scale of $\sim$ 
1000 s during the class $\alpha$.

\begin{figure}
\vskip 8.58cm
\includegraphics{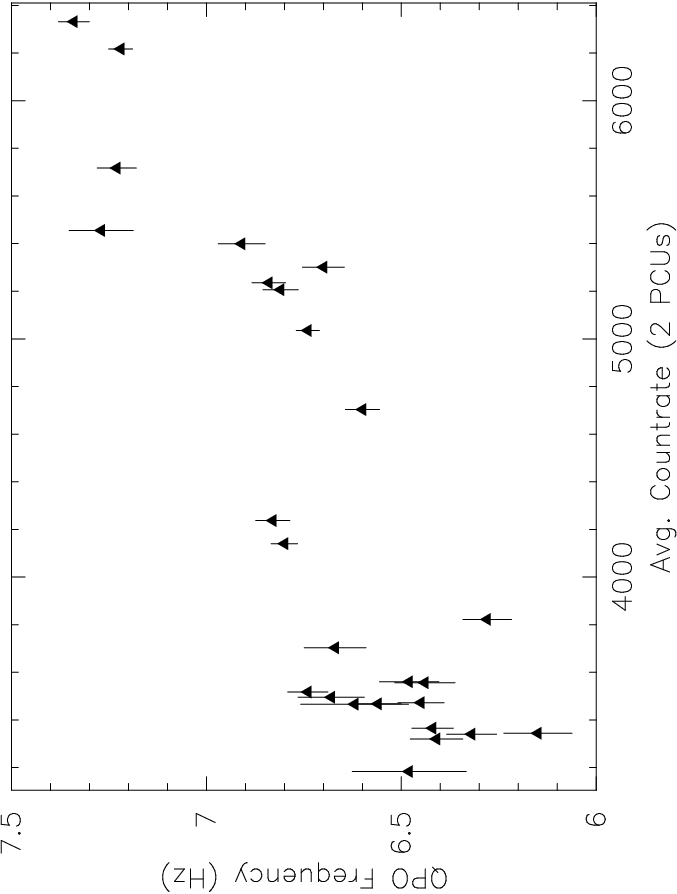}
\caption[fig4.eps]{The variation of QPO frequency with the average X-ray
count rate (for 2 PCUs) during the class $\rho$ in 2000 May$-$June from
RXTE/PCA observations is shown. A direct correlation between the average
QPO frequency with the average X-ray luminosity during the class $\rho$
is apparent in the figure.}
\end{figure}

\subsection{Spectral Analysis}

To compare the results with earlier observations of the classes $\rho$, 
$\alpha$ and $\chi2$/$\chi4$, we have selected RXTE pointed observations 
of each class when all 5 PCUs were on. The observations made on 1996 November 
19 (20402-01-03-00) and 1997 May 18 (20402-01-28-00) are used for classes 
$\rho$ and $\alpha$ respectively whereas the observation of 1996 November 
28 (20402-01-04-00) is used for the class $\chi2$. Details of these 
observations are given in Table 2. To study the spectral behaviour of the 
source during different phases, we have selected data at suitable time ranges 
for the burst/quiescent phases of the source in classes $\rho$ and $\alpha$. 
For the class $\rho$, we have selected data for the burst phase when the 
source count rate was $\geq$ 15000 counts s$^{-1}$ and for the quiescent 
phase when the source count rate was $\leq$ 10000 counts s$^{-1}$. For the 
class $\alpha$, the burst phase corresponds to the source count rate $\geq$ 
12000 and the quiescent phase to a count rate in the range of 7500 to 9500 
(for 5 PCUs). To compare the spectrum of the source during the long stretch 
($\sim$ 1000 s) of low-hard state in class $\alpha$ with that of class 
$\chi2$, we have selected the data when the source count rate was $\leq$ 
6000. To get a complete spectrum in the broad energy band (3$-$150 keV), we
have combined the data obtained from RXTE/PCA and HEXTE. Energy
spectra in 129 channel were generated from the Standard 2 mode PCA data. 
Standard procedures were applied for the data selection, background estimation
and response matrix generation (Rao et al. 2000). Systematic error of 2\% have 
been added to the PCA spectral data. For HEXTE, standard 2 as well as the 
archive mode data from the HEXTE Cluster 0, which has better spectral response,
have been used.

\begin{figure}
\vskip 6.6cm
\includegraphics{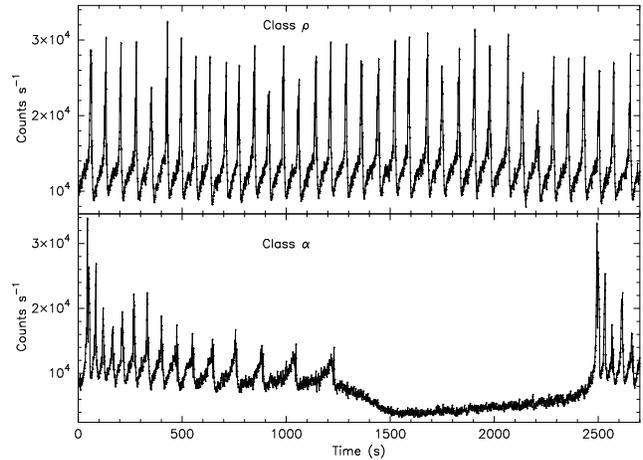}
\caption[fig6.eps]{The light curve of GRS~1915$+$105 obtained from the
RXTE/PCA data on MJD 50406 (class $\rho$) in 2 $-$ 60 keV energy range
for 1 s time bin is shown in the upper panel. The lower panel shows the
light curve of the source on MJD 50586 (class $\alpha$) in the same energy
range and time bin. The presence of regular bursts is clearly seen in the upper
panel and bursts with increasing duration are seen in the lower panel.}
\end{figure}

\begin{figure*}
\vskip 9.0cm
\includegraphics{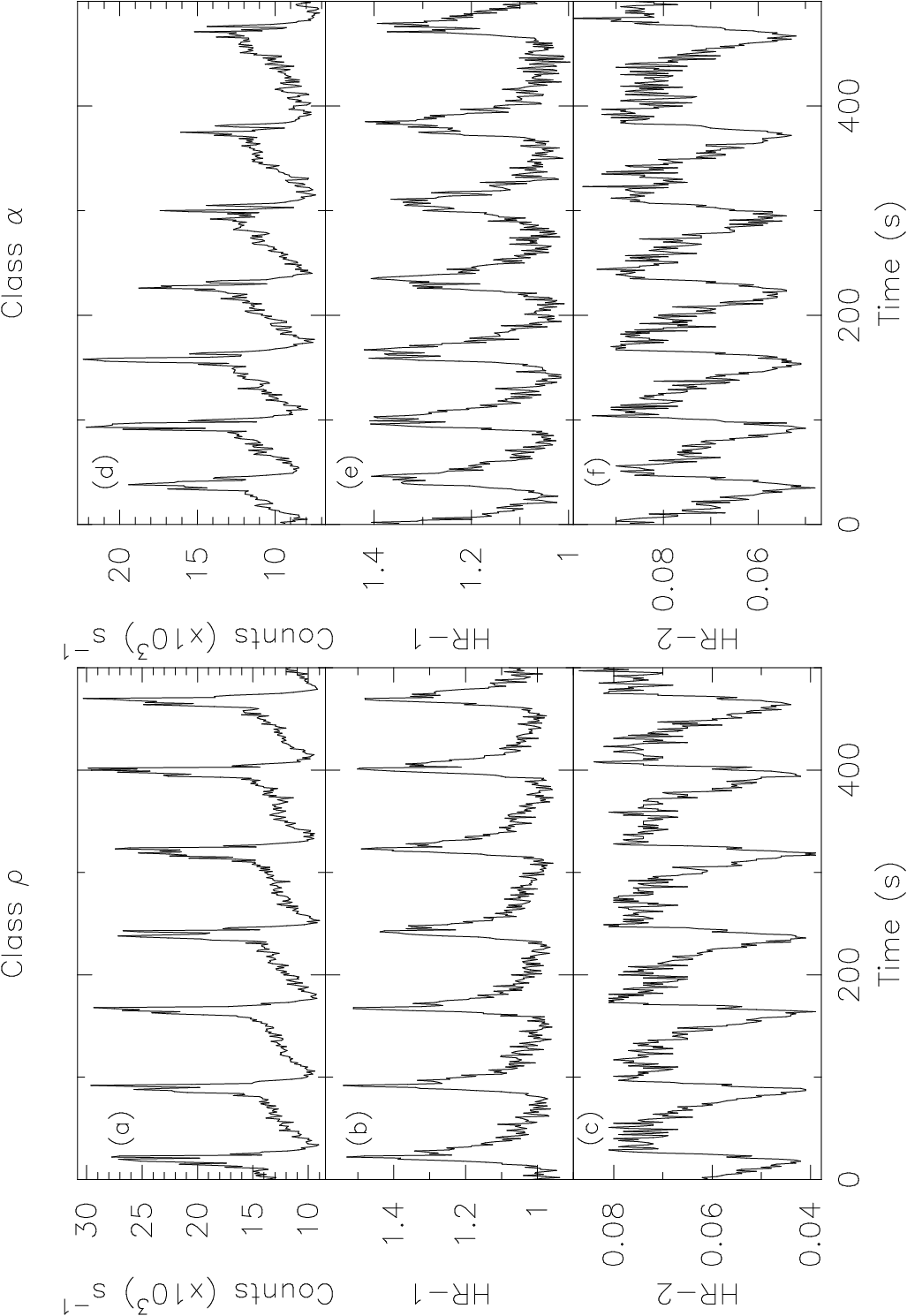}
\caption[fig7.eps]{The X-ray light curves of GRS~1915$+$105
in the 2 $-$ 60 keV range for classes $\rho$ and $\alpha$ are shown
(panels a and d respectively) along with plots of hardness ratios HR1
(count rate in 5 $-$ 13 keV / count rate in 2 $-$ 5 keV energy range) in
panels (b) and (e) and HR2 (count rate in 13 $-$ 60 keV / count rate in
2 $-$ 13 keV energy range) vs time in panels (c) and (f) respectively. From
the figure, it is seen that the burst properties in class $\alpha$
are similar to those seen in class $\rho$.}
\end{figure*}

We have fitted the spectrum for the burst and the quiescent
phases of both the classes $\rho$ and $\alpha$ using the standard black
hole model (Muno et al. 1999) consisting of ``disk-blackbody and power-law''
and ``disk-blackbody and a thermal-Compton spectrum'' with a fixed value of
absorption by intervening cold material parameterized as equivalent
Hydrogen column density N$_H$ at 6 $\times$ 10$^{22}$ cm$^{-2}$.
Simultaneous spectral fits to the PCA (in the energy range of 3$-$25 keV) 
and HEXTE (in the energy range of 15$-$150 keV) spectra were performed for 
different phases of classes $\rho$ and $\alpha$, keeping the relative 
normalization as a free parameter. The same procedure was followed for the 
fitting of the spectrum during the long stretched low-hard state of class 
$\alpha$ and the observation of class $\chi2$. The fitted parameters for 
the above two models are given in Table 3. From the table, it is observed 
that for the ``disk-blackbody and power-law'' model, there is no significant 
difference in the spectral parameters for the two different classes. For the 
``disk-blackbody and thermal-Compton spectrum'' model, it is seen that the 
disk temperatures, and the optical depth of the Compton cloud $\tau$ are 
identical for the classes $\alpha$ and $\rho$ during the different states. 
Comparing these two models, it is found that the model with disk-blackbody 
and thermal-Compton spectrum fits better than the model with disk-blackbody 
and power-law.

The resultant deconvolved energy spectra for the ``disk blackbody and 
thermal-Compton'' model are shown in Figure 8 for different phases of 
the classes $\alpha$ and $\rho$ and the long stretched low-hard states 
of classes $\alpha$ and $\chi2$. From the figure and Table 3, it is clear 
that the source spectrum is identical during  the burst and quiescent 
phases of the two X-ray classes $\alpha$ and $\rho$.

\subsection{ ASM Light Curve}

We have examined the X-ray light curves of 562 publicly available RXTE/PCA 
observations till 2000 October 11. We found that 70 observations are of 
class $\rho$ and 30 observations are of class $\alpha$. Naik and Rao (2000) 
have described the similarities in the radio properties of the source during 
both the $\rho$ and $\alpha$ classes. To examine the 
pattern of change of state of the source between the X-ray classes 
$\rho$ and $\alpha$, we show in Figure 9 the one day averaged ASM light curve 
of GRS~1915$+$105 with the identified $\rho$ and $\alpha$ classes marked
with filled circles and filled triangles respectively. 
From the figure, it is noticed that the change of state of the
source occurs from the radio-quiet low-hard state ($\chi2$/$\chi4$) to
high-state through the classes $\alpha$ and $\rho$. It is also observed that
these two classes appear together during the switching of state from
low-hard state to high state which strengthen the argument that the classes
$\alpha$ and $\rho$ have strong similarity. 
During the switching of the source from a low-hard state ($\chi2$/$\chi4$)
to the high-state, it goes through the class $\alpha$ followed by class
$\rho$ which is very clear during the transition from a long stretch of
low hard state of class $\chi2$/$\chi4$ to high state during MJD 50490 $-$
50600. Direct switching to the class $\rho$ from the low-hard state is 
not observed.

\begin{figure*}
\vskip 10cm
\includegraphics{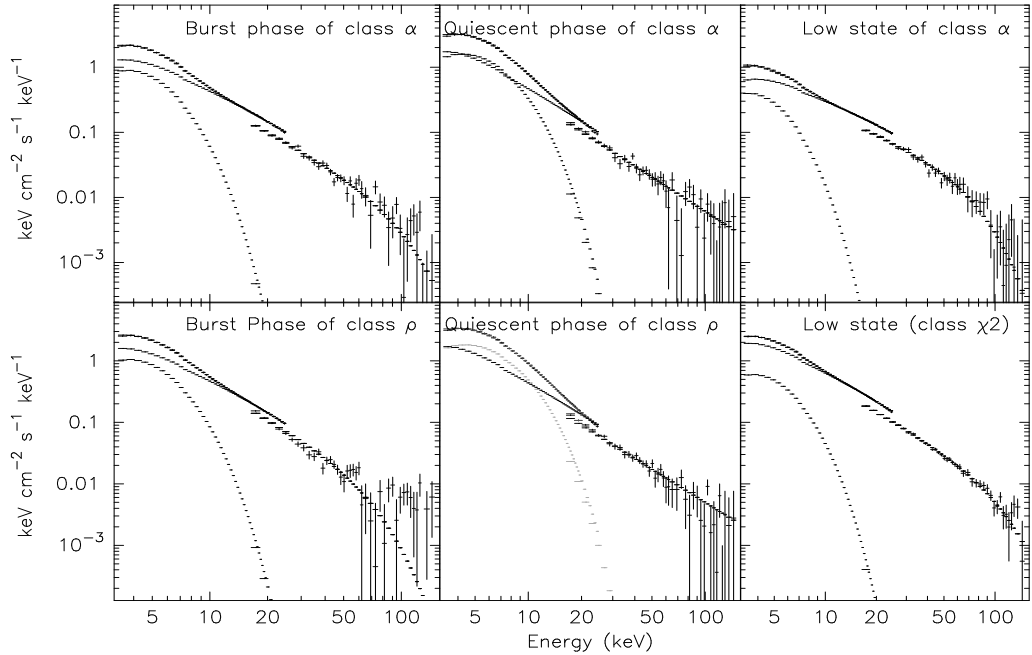}
\caption[fig7.eps]{The deconvolved X-ray spectra of GRS~1915$+$105 during
the burst and quiescent phases of the classes $\alpha$ and $\rho$ and the
low-hard states of classes $\alpha$ and $\chi2$. The fitted model consists
of a disk blackbody and thermal Compton components.}
\end{figure*}

        Based on the observed X-ray and radio properties, we find that the
X-ray classes $\rho$ and $\alpha$ are identical except for the presence of a
long stretch of low-hard state in class $\alpha$ similar to the radio-quiet
low-hard states of class $\chi2$/$\chi4$. In other-words, we can
say that the X-ray class $\alpha$ is the same as the combined class of
$\rho$ and $\chi2$/$\chi4$.

\section{Discussion}
The Galactic microquasar GRS~1915$+$105 shows extended
low-hard states on several occasions. These low-hard states are characterized
by a low frequency QPO at $\sim$ 3 Hz. The source switches from an 
extended low-hard state into a high-soft state in a wide range of time 
scales. Rao et al. (2000) have observed a slow transition from a low-hard 
state to a high-soft state (in about 3 months) in 1997 March$-$August. They 
have also reported a state transition in a very fast timescale (a few seconds) 
when the source was exhibiting irregular bursts in the X-ray light curves. 
However, the observed state transition of the source over a wide range of time 
interval is not yet clearly understood. On 2000 June 27, the source made a 
transition from a regular oscillatory behaviour (class $\rho$, in which the 
light curve contains regular bursts with recurrence time in the range 40$-$120 
s) to a steady emission (low-hard state) within 1.5 hour. The 
characteristic properties of slow rise and fast decay of these regular bursts
of class $\rho$ are unique. Taam et al. (1997) have attempted to describe 
these bursts in the framework of thermal/viscous instabilities in the 
accretion disk. Vilhu and Nevalainen (1998) tried to explain the properties
of these regular bursts by using a two phase self-consistent thermal radiative
model. Paul et al. (1998) interpreted the observed slow rise and fast decay 
of the regular bursts as the evidence for the disappearance
of matter into the event horizon of the black hole. Chakrabarti et al. (2000) 
interpreted the observed features in GRS~1915$+$105 in the light of advective 
disk paradigm which includes self-consistent formulation of shocks and 
out-flows from post-shock region and described the oscillatory behaviour 
during the X-ray classes $\rho$ and $\alpha$ as the combination of the high 
count rate ``On-state'' and the low-count rate ``Off-state''. Using this 
model, we have tried to explain the observed correlation between the QPO 
frequency and the X-ray intensity during the observations of class $\rho$.

\begin{figure}
\vskip 7.8cm
\includegraphics{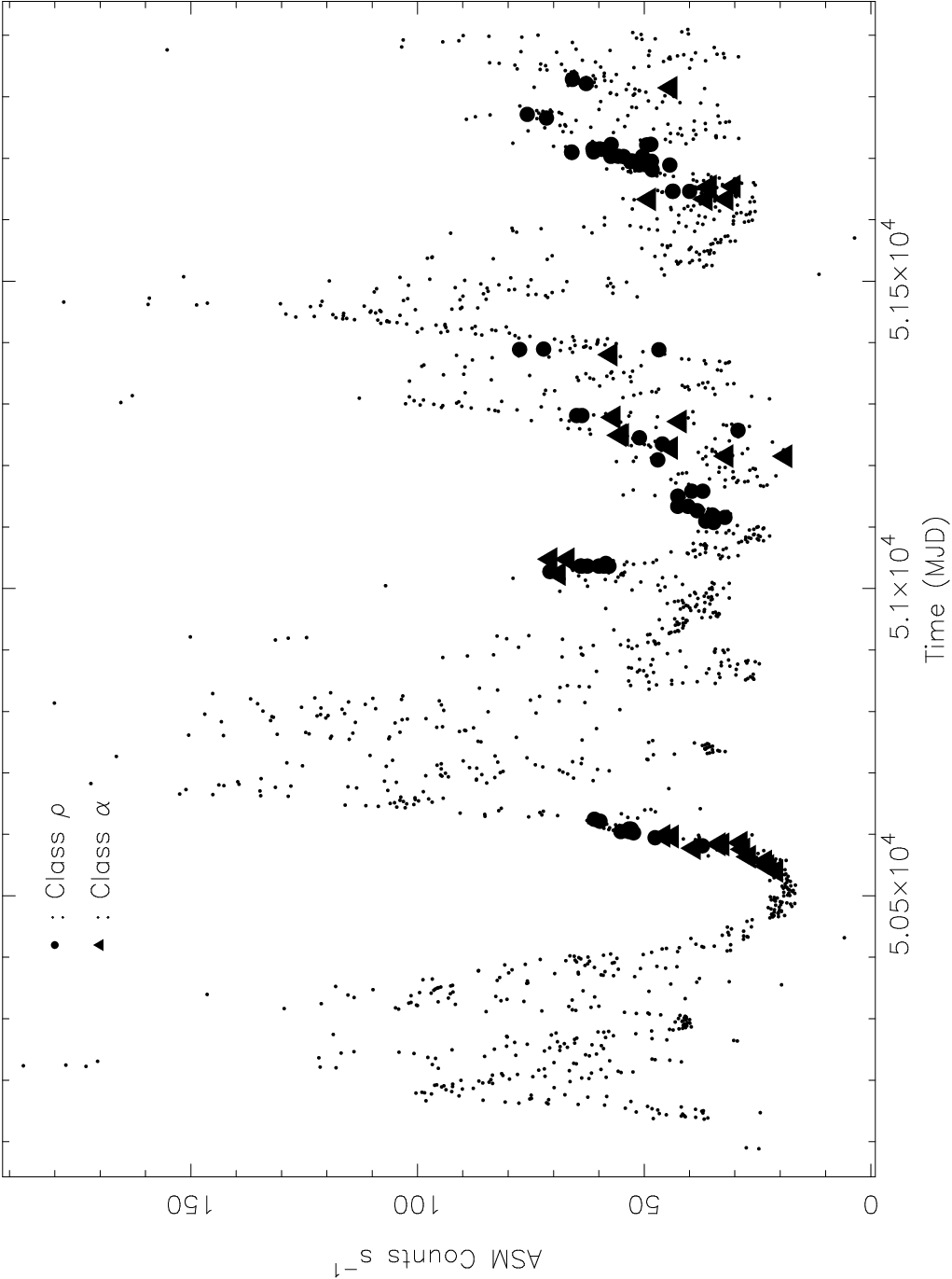}
\caption{The ASM light curve of GRS~1915$+$105 obtained from
RXTE/ASM from one day averaged dwell data. The presence of X-ray classes
$\rho$ and $\alpha$ are indicated in the figure by filled circles and filled
triangles respectively.}
\end{figure}

\begin{table*}
\centering
\caption{Spectral parameters during classes $\alpha$, $\rho$ and $\chi2/\chi4$}
\begin{tabular}{lllllll}
\hline
\hline
Parameters$^1$  &\multicolumn {2}{|c|}{Quiescent phase~~~~~~~} &\multicolumn {2}{|c|}{Burst phase~~~~~~~~~~~~~} &\multicolumn{2}{|c|}{Low-hard state~~~~~~~~}\\
  &$\alpha$ &$\rho$ &$\alpha$ &$\rho$ &$\alpha$ &$\chi2$ \\
\hline
\hline
\multicolumn {7}{|c|}{Model: Disk blackbody (diskbb) + power-law} \\
\hline
\hline
Reduced $\chi^{2}$ &1.49        &1.97     &1.49 &1.34   &1.87    &2.79\\
$\Gamma_{x}$ &2.71$\pm$0.02     &2.86$\pm$0.02  &2.75$\pm$ 0.03 &2.80 $\pm$ 0.01&2.39$\pm$0.02 &2.76$\pm$0.01 \\
$kT_{in}$ (keV)         &1.31$\pm$0.02  &1.44$\pm$0.02  &1.8 $\pm$ 0.01
&1.99 $\pm$ 0.01 &1.05$\pm$0.04 &1.31$\pm$0.2\\
HEXTE Count rate        &63$\pm$1       &63$\pm$1       &68$\pm$1
&62$\pm$1       &61$\pm$1 &93$\pm$1\\
PCA Count rate          &8001$\pm$37    &9474$\pm$45    &12647$\pm$59
&13994$\pm$64   &4336$\pm$19 &9973$\pm$44 \\
\hline
\hline
\multicolumn {7}{|c|}{Model: Disk blackbody (diskbb) + thermal-Compton (CompST)} \\
\hline
\hline
Reduced $\chi^{2}$      &0.99     &1.53     &1.46     &1.30        &0.79 &0.9\\
$kT_{in}$ (keV) &1.32 $\pm$ 0.02  &1.37 $\pm$ 0.02  &1.79 $\pm$ 0.02
&1.96 $\pm$ 0.01 &1.17 $\pm$ 0.03   &1.35 $\pm$ 0.03\\
$kT_{e}$ (keV)  &19.45 $\pm$ 2.1  &13.14 $\pm$ 1.28 &57.45 $\pm$ 28.4
&30.39 $\pm$ 7.85 &15.73 $\pm$ 1.08  &21.82 $\pm$ 1.2 \\
$\tau$          &2.87 $\pm$ 0.23  &3.49 $\pm$ 0.26  &1.24 $\pm$ 0.49
&1.96 $\pm$ 0.37   &3.91 $\pm$ 0.2 &2.66 $\pm$ 0.1\\
\hline
\hline
\end{tabular}
\begin{flushleft}
$^1$ : $\Gamma_{x}$ : Power-law photon index, kT$_{in}$ : Inner disk
temperature, kT$_{e}$ : Temperature of the Compton cloud, $\tau$ :
Optical depth of the Compton. \\
\end{flushleft}
\end{table*}

Chakrabarti \& Manickam (2000) derived a relation between the QPO
frequency in 1$-$10 Hz range and duration of the quiescent phases
(class $\rho$ and $\alpha$). As the 1$-$10 Hz QPO could be due to the
oscillation of shocks located at tens to hundreds of Schwarzschild
radius, they tried to explain the switching of burst phase to quiescent 
phase and vice versa and the duration of the quiescent phase by assuming
an outflow from the post-shock region. Assuming the flow to be isothermal
until R$_c$ (the location of the sonic point), a shock compressed gas
with R (the shock compression ratio) $>$ 1, must produce outflows which
pass through the sonic points. Hence matter is subsonic until the sonic
point. As the Compton cooling becomes catastrophic when the Thomson
scattering opacity (k$_{es}$) becomes 0.4, the duration of the off
state (i.e. the duration between the end of a burst and the beginning
of the next burst which we described as the quiescent phase) is given
by t$_{off}$ $=$ (4$\pi$R$^{2}_{c}$)/(3 $\dot{M}_{out}$ k$_{es}$).
Using the relation between inflow and outflow rates (Chakrabarti 1998),
$\alpha$ = 3/2 (for a low angular momentum freely falling matter),
and v$_0$ = 1, the duration of the off state is given by

\begin{equation}
t_{off}=461.5~(\frac{0.1}{\Theta_{\dot{M}}})~(\frac{M}{10 M_{\odot}})^{-1/3}~ \nu^{-4/3}~ s
\end{equation}
where 
\begin{equation}
\Theta_{\dot{M}} = \frac{\Theta_{out}}{\Theta_{in}} \frac{\dot{M}}{\dot{M}_{Edd}}
\end{equation} 
$\Theta_{in}$ and $\Theta_{out}$ are the solid
angles of the inflow and the outflow respectively.
Keeping the numerical coefficient constant in the above
equation by putting the value of $\alpha$ = 1 and v$_0$ = 0.066,
which is very reasonable for a black hole accretion, one obtains
 
\begin{equation}
t_{off}=461.5~(\frac{0.1}{\Theta_{\dot{M}}})~(\frac{M}{10 M_{\odot}})^{-1}~(\frac{v_0}{0.066})^2~ \nu^{-2}~s
\end{equation}

This relation has been found to be valid for all the observations.
According to this relation, the greater is the value of the duration of the
off state t$_{off}$, less is the frequency of the QPOs. If one compares
the presence of QPOs and the duration of the off state (quiescent phase) in
the Figure 1(d)  of Muno et al. (1999), it can be seen that, during class
$\rho$ the QPOs are at similar frequencies.
Also in their Figure 1(j), during the on and off states of class 
$\alpha$, the QPO frequencies change according to the above relation. 

In class $\rho$, the relation between QPO frequency and t$_{off}$ should 
be valid for a given day and the proportionality constant varies for each 
day. We find that t$_{off}$$\nu$$^2$ is correlated with the quiescent
count rate with a correlation coefficient of 0.94. This can be explained
by the fact that for a given day $\Theta_{\dot{M}}$ can change (see eqn. 1)
and it should be related to the quiescent count rate.

Although different states of the microquasar GRS~1915$+$105 are explained
by various models, none of the models explain the transition from one
state to other. It would be
interesting to study the changes in the emission processes in the
accretion disk and the factors which trigger the change of state as well
as the transition from one class to another in the source.

\section*{Acknowledgments  }

We acknowledge the contributions of the scientific and technical staff
of TIFR, ISAC and ISTRAC for the successful fabrication, launch and
operation of the IXAE. We thank Dr. S. Seetha and Dr. K. Kasturirangan
for their contribution to the IXAE. It is a pleasure to acknowledge constant  
support of Shri K. Thyagarajan, Project Director IRS$-$P3 satellite, Shri 
R. N. Tyagi, Project Director IRS$-$P4 satellite, Shri
J. D. Rao and his team at ISTRAC, Dr P. S. Goel, Director ISAC and the 
Director of the ISTRAC. SN thanks M. Choudhury for his comments and discussions 
on the manuscript. This research has made use of data obtained through 
the High Energy Astrophysics Science Archive Research Center Online Service, 
provided by the NASA/Goddard Space Flight Center. Finally we thank an anonymous
referee for his comments and suggestions which resulted in very significant 
improvement of this paper.

\end{document}